\def\year{2021}\relax
\documentclass[letterpaper]{article} 
\usepackage{lmodern}
\usepackage{aaai21}  
\usepackage{times}  
\usepackage{helvet} 
\usepackage{courier}  
\usepackage[hyphens]{url}  
\usepackage{graphicx} 
\usepackage{natbib}  
\usepackage{caption} 
\usepackage{amsfonts}
\usepackage{multirow}
\usepackage{booktabs}
\title{Learning compositional programs with arguments and sampling}
\author {
    Giovanni De Toni\textsuperscript{\rm 1},
    Luca Erculiani\textsuperscript{\rm 1},
    Andrea Passerini\textsuperscript{\rm 1} \\
}
\affiliations {
    \textsuperscript{\rm 1} Structured Machine Learning Group (SML),\\
    Department of Information Engineering and Computer Science (DISI)\\
    University of Trento, Italy\\
    \texttt{\{giovanni.detoni, luca.erculiani, andrea.passerini\}@unitn.it}
}

\usepackage{amsmath}
\usepackage{subfig}
\graphicspath{ {./images/} }
\DeclareMathOperator*{\argmax}{arg\;max}

\ifdefined\releaseversion
    \newcommand{\andrea}[1]{}
    \newcommand{\luca}[1]{}
    \newcommand{\giovanni}[1]{}
\else
    \usepackage[dvipsnames]{xcolor}
    \newcommand{\andrea}[1]{{\bf \textcolor{blue}{Andrea: #1}}}
    \newcommand{\luca}[1]{{\bf \textcolor{cyan}{Luca: #1}}}
    \newcommand{\giovanni}[1]{{\bf \textcolor{red}{Giovanni: #1}}}
\fi

\begin{document}

\maketitle

\begin{abstract}
One of the most challenging goals in designing intelligent systems is empowering them with the ability to synthesize programs from data. Namely, given specific requirements in the form of input/output pairs, the goal is to train a machine learning model to discover a program that satisfies those requirements.
A recent class of methods exploits combinatorial search procedures and deep learning to learn compositional programs. However, they usually generate only toy programs using a domain-specific language that does not provide any high-level feature, such as function arguments, which reduces their applicability in real-world settings.
We extend upon a state of the art model, AlphaNPI, by learning to generate functions that can accept arguments. This improvement will enable us to move closer to real computer programs. Moreover, we investigate employing an Approximate version of Monte Carlo Tree Search (A-MCTS) to speed up convergence. We showcase the potential of our approach by learning the Quicksort algorithm, showing how the ability to deal with arguments is crucial for learning and generalization.
\end{abstract}

\section{Introduction}

The ability to autonomously synthesize programs from data is one of the main goals in developing intelligent systems. 
Namely, given specific requirements, such as input/output pairs or more formal specifications, the model must learn a program that meets those demands. Several approaches to try to solve this task are based on the neural controller - interface framework, in which a neural network interacts with an external structured environment by using primitives, such as read/write heads or other atomic functions~\cite{pmlr-v48-zaremba16, reed2016neural, DBLP:journals/corr/GravesWD14, graves2016hybrid, pierrot2019learning, pierrot2020learning}.
Neural Turing Machines \cite{DBLP:journals/corr/GravesWD14}, and Differentiable Neural Computers \cite{graves2016hybrid} can learn simple procedures by training on input/output pairs. These models use external memory, and they are differentiable, which means that they can be trained end-to-end by gradient descent. However, they are essentially black boxes since it is non-trivial to inspect them to understand their discovered algorithm. More importantly, supervision is provided at the level of full program outputs only, making training strongly susceptible to overfitting as shown for the priority sort task of \citet{DBLP:journals/corr/GravesWD14}.

Neural-Programmer interpreters (NPI) \cite{reed2016neural} use a reinforcement learning paradigm to learn highly compositional programs by exploiting execution traces. An execution trace is an ordered list of commands that, if executed, solve a given task. Execution traces provide finer-grain supervision during training, encouraging convergence towards more accurate programs. NPI assumes to have at hand a large dataset of execution traces for the given problem, substantially limiting its applicability in real-world scenarios, where an oracle capable of providing such a dataset could be unavailable. A recent model, named AlphaNPI, shows how to train NPI without execution traces. AlphaNPI can solve both discrete and continuous control problems (\citet{pierrot2019learning, pierrot2020learning}), and it works by combining NPI with Monte Carlo Tree Search (MCTS), which is used to discover the execution traces. This approach showed outstanding results, both in terms of generalization and interpretability of the final models. However, the spectrum of tasks it can learn is still somewhat limited since it generates programs that use an ad-hoc language that does not present any high-level construct of a formal programming language such as loops, variables or function arguments.

We can also induce programs from input/output examples by creating logic programs. Inductive Linear Programming (ILP) \citet{MUGGLETON1994629, ilp2020} is a classical framework to learn logic programs from data. ILP is data-efficient and also supports \textit{predicate invention} through the use of meta-rules. Namely, ILP systems can invent new programs and functions. However, these techniques require strong formalization of the problem settings into logical form. They require extensive domain knowledge, and it can be hard to provide a formalization to learn more complex real-world programs. Moreover, they are also sensitive to noisy programs since by learning underperforming procedures, we can undermine the performances of the final learned algorithm.

This work presents some promising results towards generalized neural interpreters and discusses additional future challenges that still need to be addressed. We also provide a brief discussion comparing ILP and neural-based models. The main contributions are the following:
\begin{itemize}
    \item We propose a new version of the AlphaNPI model, which brings us closer to learning real-world programs by learning functions that also accept arguments.
    \item We propose an Approximate recursive Monte Carlo Tree Search (A-MCTS) procedure to improve convergence.
    \item We developed a refined training strategy featuring re-training over failed tasks that improves robustness and generalization.
\end{itemize}
\begin{figure}
    \centering
    \includegraphics[width=0.7\linewidth]{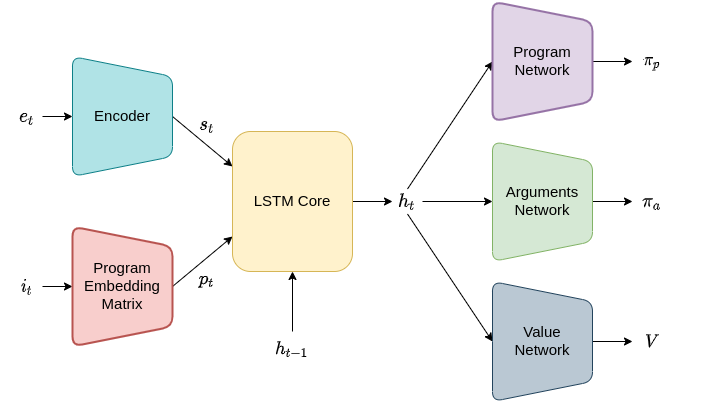}
    \caption{Complete Diagram of the Architecture}
    \label{fig:architecture}
\end{figure}

\section{Problem Statement and Original Setting}

We consider the problem of learning a complex algorithm, such as Quicksort, by having an agent interacting with an environment $e$ by choosing actions $p$. The initial actions are called \textit{atomic actions}.
In the Quicksort case, the environment is a list of integers, and the atomic actions are element swaps and one-step pointer moves. The agent learns to leverage and to combine these atomic actions to produce higher-level programs. These learnt programs are then added to the collection of available actions. Therefore, the agent can benefit from both atomic and more complex actions to produce advanced algorithmic behaviour.
Each action has assigned a \textit{level}. The level 0 actions are the atomic actions that do not need to be learned. Discovered programs have positive levels. Each program can only call lower-level action or lower-level learned programs. The original work supported also recursion, by letting programs calling themselves. However, in this work, we do not experiment with recursive programs.
In addition, each program and atomic action has associated \textit{pre-conditions}, in the form of boolean constraints, conditioned on the environment state. Pre-conditions ensures we are consistently generating programs that are correct and feasible. For instance, the pre-condition of the Quicksort program is that all pointers must be at the beginning of the list. Pre-conditions are not learned, but they are provided beforehand. The objective is to learn a library of programs organized into a hierarchy. In our case, each task corresponds to learning a single program (e.g., the \texttt{partition} function of Quicksort). The reward function returns 1 if the program is correct, 0 otherwise. The goal of the agent is to maximize the expected reward for all the tasks.
This goal configures as a multi-task reinforcement learning and as a discrete search problem. The search space is sparse since there exist many possible programs but very few viable ones. It also makes the reward function sparse since we get a positive reward and thus learn if and only if we obtain the correct program.

\section{Novel Setting}

We change the original setting by making actions and programs accept \textit{arguments}, $a$. Each action/program can take an ordered list of at most three arguments. In our environment, these arguments are the pointers used to manipulate the list. For example, the \texttt{quicksort} program accepts two arguments, the pointer placed at the beginning of the list and the pointer placed at the end of the list. Additionally, an action/program can also accept no arguments, such as the \texttt{stop} action, which terminates the execution of the simulation. As for the pre-conditions, the maximum number of arguments that a program can accept is given apriori.
We also extend the pre-conditions to consider the arguments to decide if a program is feasible. For example, the \texttt{swap} pointer action cannot be called by giving as argument the same pointer twice.

\subsection{Architecture}

We start from the architecture developed by~\citet{pierrot2019learning}, and adapt it to the novel setting by introducing an additional module, $f_{arguments}$. Thanks to this new module, we learn compact programs that can adapt their behaviour by accepting arguments instead of learning a separate program for each possible value of the arguments. For example, we can learn a \texttt{move} action to shift pointers that can operate on multiple pointers at the same time, \texttt{move(pointer\_1,pointer\_3)}. With the original method, we would have had to learn two separate actions, \texttt{move\_pointer\_1} and \texttt{move\_pointer\_3}.

The architecture is shown in Figure~\ref{fig:architecture} and consists of six modules: an encoder ($f_{enc}$), a programs matrix ($M_{prog}$), an LSTM core ($f_{lstm}$), a program network ($f_{program}$), an arguments network ($f_{arguments}$) and a value network ($f_{value}$). The encoder takes as input an environment $e_t$ and encodes it as a set of features $s_t$. The program matrix takes as input an index $i_t$ and returns the corresponding program embedding $p_t$. The LSTM core executes programs while conditioning on $p_t$, the feature set $s_t$ and its internal state $h_{t-1}$. The program and arguments  networks take the LSTM output, $h_t$, and return probabilities over the action space, $ \pi_{p}$, and argument space, $\pi_{a}$, respectively. Lastly, the value network takes instead the LSTM output to estimate the value function $V$.
The equation of the architecture are the following:
\begin{align}
    & s_t = f_{enc}(e_t) \qquad p_t = M_{prog}[i_t]\\
    & h_t = f_{lstm}(s_t, p_t, h_{t-1}) \\
    & \pi_{p} = f_{program}(h_t) \quad \pi_{a} = f_{arguments}(h_t) \\
    & V = f_{value}(h_t)
\end{align}
\begin{figure*}[th!]
    \centering
    \includegraphics[width=0.7\linewidth]{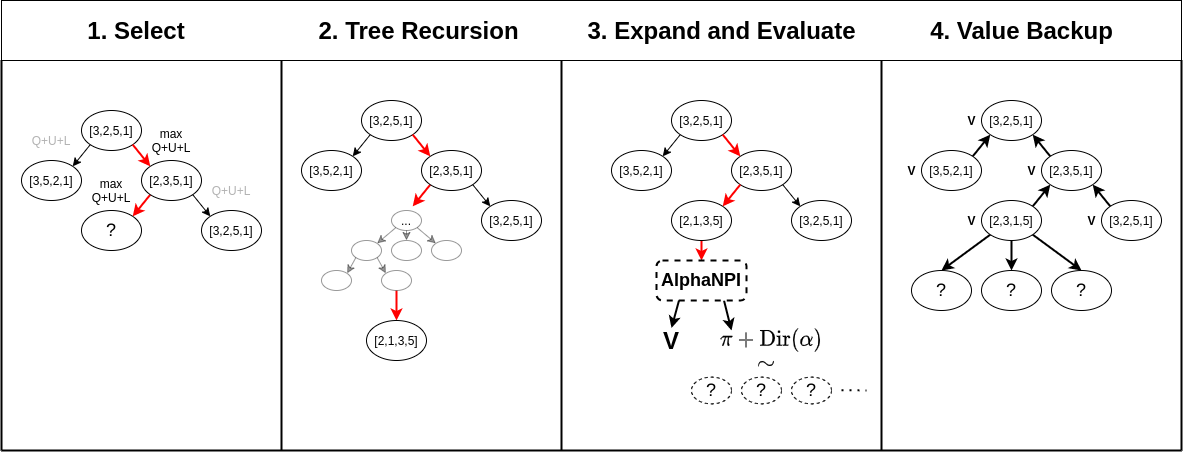}
    \caption[Approximate Monte Carlo Tree search.]{1. Each simulation explores the tree by selecting those actions that maximize a target objective. 2. If we find a new node that is not an atomic function, we run A-MCTS recursively by resetting to zero the LSTM state. 3. Given the expanded leaf node, we compute the program and actions policies ($\pi_p^{mcts}$ and $\pi_a^{mcts}$). We sample from the two policies to add $n$ new child nodes ($n << M$ where $M$ is the total possible child nodes ). 4. We return the value V, and we propagate it back in the tree.}
    \label{fig:MCTS}
\end{figure*}

\subsection{Approximate Monte Carlo Tree Search}

Monte Carlo Tree Search \cite{coulom2006backupmonte, Kocsis06banditbased} is an algorithm to search large combinatorial spaces represented by trees efficiently. It was successfully used to solve many tasks, such as mastering complex games and protein folding \cite{silver2017mastering, AlphaFold2021}. The original method is made by four phases: \textit{selection}, \textit{recursion}, \textit{expansion} and \textit{backup}. Many different flavours of MCTS are available (for instance, \citet{DBLP:journals/corr/abs-2103-04931, pmlr-v119-grill20a, xiao2018memory}). In this work, we build upon the recursive MCTS developed by \citet{pierrot2019learning}. We propose an Approximate MCTS (A-MCTS) by extending the \textit{expansion} phase not to expand all the possible nodes available.
This improvement is needed to account for larger search spaces that would be prohibitive to explore thoroughly. In our setting, we have discrete arguments, and the original MCTS requires creating a node for each program/argument pair available. If we increase the number of available arguments, the number of program/argument combinations grows significantly.

In our setting, each node of the tree represents the environment's state at time \textit{t} and each transition represents a \textit{program call}. Figure \ref{fig:MCTS} shows a description of the A-MCTS algorithm. Given the policies $\pi_p^{mcts}$ and $\pi_a^{mcts}$, we add Dirichlet noise to foster exploration and we sample $n$ program/arguments pairs, $(p_i, a_i)$, ($n$ is a hyperparameter of the model) and we add them as possible future available actions. The simulation phase will be shorter with fewer nodes. By exploring fewer nodes, we reward immediately good solutions. Thanks to the random perturbations, we will eventually explore all the available configurations if good solutions are sparse.

\section{Training}

During a training iteration, the agent selects a program $p_t$ to learn. It executes $n_{ep}$ episodes by exploring the search space with A-MCTS. The data gathered during the episodes are aggregated to construct the tree policy vectors for both programs and arguments, $\pi_p^{mcts}$ and $\pi_a^{mcts}$. After each successful episode, we collect the observations, the hidden states, the task indexes, the rewards, $\pi_p^{mcts}$ and $\pi_a^{mcts}$ into an execution trace. More formally, an execution trace for a given program $p_t$ is a tuple $(e_I, p_t, h_t, e_O, r_t, \pi_p^{mcts}, \pi_a^{mcts})$. An execution trace gives the exact sequence of actions that, if applied to the input $e_I$, produce the output $e_O$. The final discovered execution traces are stored in a replay buffer. Then, a mini-batch is sampled from the replay buffer, and the agent is trained on this data to minimize this adjusted loss function:

\begin{equation}
    l = \sum_{batch} -\underbrace{(\pi^{mcts}_p)^T\log{\pi_p}}_{l_{policy}} -\underbrace{(\pi^{mcts}_a)^T\log{\pi_a}}_{l_{arguments}} + \underbrace{(V-r)^2}_{l_{value}}
\end{equation}

We want to push the network to reproduce the execution traces found by jointly minimizing the cross-entropy between the policies discovered by MCTS and the policies generated by the network.
$l_{policy}$ minimizes the cross-entropy between the program policies, $\pi_{p}^{mcts}$ and $\pi_{p}$. $l_{arguments}$ minimizes the same cross-entropy but between the two arguments policies, $\pi_{a}^{mcts}$ and $\pi_{a}$. Instead, $l_{value}$ pushes the network to generate the correct value function $V$.

We also employ curriculum learning \cite{pmlr-v70-andreas17a} to focus on learning programs by following the hierarchy, such as to learn lower-level programs first. Moreover, curriculum learning ensures choosing the next program to learn by looking at the success rate of the single programs. This rule means that programs that fail often are picked more often for learning.

Given the environment state, the agent emits a program policy $\pi_p$ and an argument policy $\pi_a$. The policies are such that $\sum_M \pi(i) = 1$, where M is the total number of available programs or arguments. The best program/arguments pair $(p^*, a^*)$ is given by:

\begin{equation}
    p^* = \argmax(\pi_p) \qquad a^* = \argmax(\pi_a)
\end{equation}


\subsection{Re-train over failed environments}

We noticed that a small fraction of training environments are substantially harder to solve and are thus mostly neglected when optimizing average performance during training. The result is that sub-optimal programs are learned, and the performance degradation becomes apparent when programs are evaluated on the longer lists used for testing. To tackle this problem and improve robustness and generalization, we implemented a function $g(p_i)$ to re-train over previously failed environments.
Namely, given a task $T_i$ to learn, if we get a 0 reward, we will store the used environment $e_I$ into a buffer called $e_{failed}$. Then, if we happen to try to learn the task $T_i$ again, we will run A-MCTS by sampling with a small probability $\epsilon$ an environment from $e_{failed}$ in which the model had failed to solve in previous iterations. In the other case, we sample the next state from a normal distribution over the environment conditioned on the task $T$.
\begin{equation}
    g(p_i) = \begin{cases}
     e_I \sim e_{failed} & \epsilon \\
     e_I \sim \mathcal{N} & 1-\epsilon
    \end{cases}
\end{equation}
The buffer $e_{failed}$ implements some curriculum learning since, given task $T_i$, if we use previously failed states, we will sample the failed states by looking at the success rate. If a state fails more often than others, then it will be sampled more frequently.

\begin{table*}[h!]
\centering
\begin{tabular}{llllll}
\toprule
                                                \textbf{Program}      & \textbf{List Length} & \textbf{$M_{alphanpi}$} & \textbf{$M_{alphanpi}^S$} & \textbf{$M_{alphaargs}$} & \textbf{$M_{alphaargs}^S$} \\ \midrule
\multicolumn{1}{c}{\texttt{partition\_update}} & 5 & \textbf{1.00} & \textbf{1.00} & \textbf{1.00} & 0.96 \\
 & 10 & \textbf{1.00} & \textbf{1.00} & \textbf{1.00} & 0.98 \\
 & 20 & \textbf{1.00} & 0.98 & \textbf{1.00} & 0.96 \\
 & 40 & \textbf{1.00} & \textbf{1.00} & \textbf{1.00} & 0.94 \\
 & 60 & \textbf{1.00} & \textbf{1.00} & \textbf{1.00} & 0.96 \\ \midrule
\multirow{5}{*}{\texttt{partition}} & 5 & \textbf{1.00} & 0.98 & \textbf{1.00} & 0.96 \\
 & 10 & \textbf{1.00} & 0.78 & 0.98 & 0.78 \\
 & 20 & \textbf{1.00} & 0.80 & \textbf{1.00} & 0.70 \\
 & 40 & \textbf{1.00} & 0.80 & \textbf{1.00} & 0.82 \\
 & 60 & \textbf{1.00} & 0.76 & 0.98 & 0.80 \\ \midrule
\multirow{5}{*}{\texttt{quicksort\_update}} & 5 & 0.72 & 0.58 & \textbf{1.00} & 0.60 \\
 & 10 & 0.64 & 0.42 & \textbf{0.96} & 0.48 \\
 & 20 & 0.56 & 0.34 & \textbf{0.90} & 0.56 \\
 & 40 & 0.36 & 0.30 & \textbf{0.98} & 0.40 \\
 & 60 & 0.38 & 0.26 & \textbf{0.94} & 0.44 \\ \midrule
\multirow{5}{*}{\texttt{quicksort}} & 5 & 0.10 & 0.02 & \textbf{1.00} & 0.06 \\
 & 10 & 0 & 0 & \textbf{0.66} & 0 \\
 & 20 & 0 & 0 & \textbf{0.42} & 0 \\
 & 40 & 0 & 0 & \textbf{0.22} & 0 \\
 & 60 & 0 & 0 & \textbf{0.02} & 0 \\ \bottomrule
\end{tabular}
\caption{Generalization accuracy of the learned programs by varying the length of the list. Each model was tested on 50 different random lists for each given length. The accuracy gives the fraction of lists correctly sorted. The bold values indicate the best results for that given list length.}
\label{tab:models-accuracy}
\end{table*}

\section{Experiments}

We tested our approach on a sorting task. It is a familiar setting used by many others \cite{reed2016neural, pierrot2019learning, DBLP:journals/corr/GravesWD14}. We focused on learning the hierarchy of programs needed to perform the Quicksort algorithm. Unlike the Bubblesort algorithm used in previous works~\cite{pierrot2019learning}, Quicksort cannot be easily written without functions that can accept arguments. Therefore, we choose it as the target for our experiments.
Quicksort is a divide-and-conquer algorithm \cite{10.1093/comjnl/5.1.10}. It works by selecting a \textit{``pivot"} value that partitions the list into sub-lists which it then sorts. The algorithm has an average time complexity of $O(n \log n)$, also presenting good space complexity, $O(\log n)$ extra bits to sort $n$ elements. The algorithm is more refined to reach these performances, especially the method used to select the pivot. 
The code for the experiments is freely available\footnote{\url{github.com/geektoni/learning_programs_with_arguments}}.

We consider an environment composed of three main components: a list of $n$ integers, three pointers that can reference the list's elements and a stack that can save and retrieve the pointers' current positions.
The programs can accept at most three arguments. Each argument represents a reference to one of the three pointers available in the environment. We also added an empty value to signal arguments that the given function has to ignore. Additionally, as in regular  programming languages, the arguments order count (e.g., \textit{function(a,b,c)} has a different meaning from \textit{function(b,c,a)}).

We trained the architecture with lists from 2 to 7 elements. The list's elements were integers, and we constrained them between 0 and 10. We trained four different models. 
$M_{alphanpi}$ is the baseline, and it uses the vanilla architecture from \citet{pierrot2019learning}. $M_{alphanpi}^{s}$ is again a baseline model which uses our A-MCTS instead. $M_{alphaargs}$ and $M_{alphaargs}^s$ employ our custom architecture presented in the previous sections. The former uses the standard MCTS, and the latter uses A-MCTS instead. 

The higher-level programs that compose our hierarchy and which will be learned are: \texttt{partition\_update}, \texttt{partition}, \texttt{quicksort\_update} and \texttt{quicksort}. They correspond to certain pieces of the original Quicksort algorithm. The complete lists and descriptions of learnable programs and atomic actions can be found in the Appendix.
The $M_{alphanpi}$ and $M_{alphanpi}^S$ models learn functions that do not accept arguments. Therefore, we provided them with an augmented set of atomic actions, which we built by taking the cartesian product between the atomic actions and the possible arguments they can accept. 

\section{Discussion}

While training, we also recorded the total number of nodes expanded by the MCTS and A-MCTS procedures after each training step. The idea is to check if we can converge to a solution by doing an approximate exploration of the search space by expanding fewer nodes. 

As further evaluation studies, we validated the trained models on lists of up to 60 elements to investigate the generalization capabilities. Table \ref{tab:models-accuracy} presents the validation results for all the learnable programs.

From the results, the $M_{alphaargs}$ is the only model which can learn the Quicksort algorithm correctly, whereas the other models fail. It also shows some generalization capacities since it can sort lists with a different length than those used for training. We can also see how the various models struggled when learning the \texttt{quicksort\_update} function. 

\begin{figure*}[h!]
\centering

\subfloat{
\includegraphics[width=0.9\linewidth]{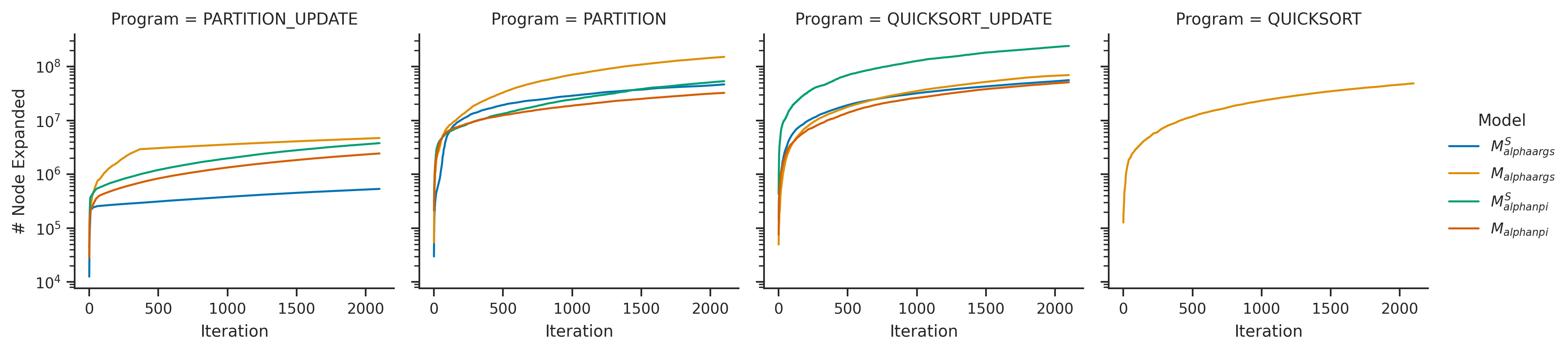}
}

\subfloat{
\includegraphics[width=0.9\linewidth]{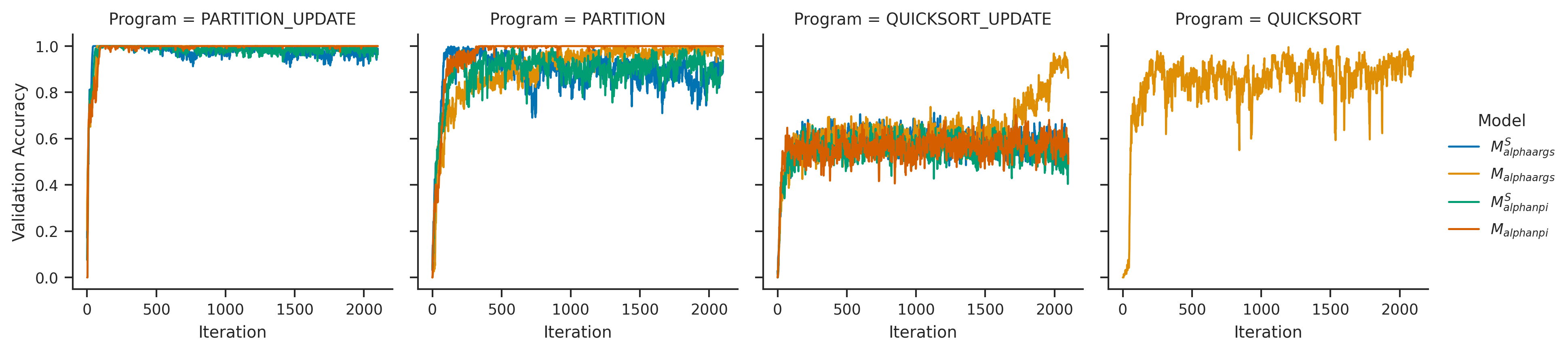}
}

\caption{Total node expanded by models (top) and the corresponding accuracy (bottom) while training. The node count is expressed with the log scale. In some cases, the Approximate MTCS shows to enable convergence by exploring fewer nodes. Moreover, once the models learn a program, the node expansion slows down. Interestingly, it seems that the \texttt{quicksort\_update} function is harder to learn. Unfortunately, only one model is able to learn the complete Quicksort procedure.}
\label{fig:model}

\end{figure*}

Interestingly, the models trained with the approximate procedure A-MTCS show similar convergence rates to those trained with MCTS. They also show good generalization performances, albeit not reaching the level of the original MCTS. From Figure \ref{fig:model}, we can look at the total nodes expanded by both the approximate and exact procedures. For specific programs, such as \texttt{partition\_update} or \texttt{partition}, the A-MCTS can converge to a solution by exploring fewer nodes with respect to the counterpart. However, the A-MCTS behaviour seems to be less stable than MCTS, and further investigations are required to be able to exploit its exploration efficiency without losing robustness.

\subsubsection{Comparison with ILP} Currently, many common program synthesis problems can be solved either by ILP or by neural-interpreter methods. We argue that neural-based solutions require less formalization and specifications to learn an algorithm. However, they are more data-intensive than a standard ILP procedure. As we move closer to real-world programs, we think that neural-based models are easier to use since they rely on looser specifications. Moreover, we think the use of a differentiable model can speed up convergence when used as a \textit{prior} for the discrete search method. In our case, a differentiable model implicitly learns a representation of the target program given the traces, thus leading to faster trace discovery.
\section{Conclusion}

This work presents some initial results to improve program synthesis by generating more human-like programs starting from input/output examples.
We propose an improved AlphaNPI architecture to learn programs that can accept arguments to diversify their behaviour. The support for arguments enables our revised architecture to produce richer programs and learn more complex algorithms with respect to the original work. It also generates code that is more similar to the one produced with current high-level programming languages. Additionally, we present an Approximate Monte Carlo Tree Search method that enables us to converge by exploring only a fraction of the search space. We benchmarked our method by learning a well-known sorting algorithm, Quicksort, showing how our approach manages to learn it effectively while AlphaNPI fails to converge. We also show how the learned program generalizes when sorting lists of increasing length.
This work is a first attempt at enriching neural program synthesis algorithms with additional features of high-level programming languages. Many open issues are still to be addressed. First, while the approximate MCTS procedure effectively reduces the search space and can potentially contribute to scale up the complexity of learnable programs, further investigations are needed to improve its robustness. Second, in this work only atomic actions can accept arguments, whereas higher-level programs can accept only the ``empty" argument. This behaviour happens because all methods operate in the same environment, and there is no proper program scope. Additional studies are required to add this notion of program space to better use the new arguments.

\begin{quote}
\begin{small}
\bibliography{./references.bib}
\end{small}
\end{quote}

\onecolumn

\section{Appendix}

Table \ref{tab:quicksort-programs} and \ref{tab:quicksort-prog-args} show the complete hierarchy of programs. The former shows programs for the $M_{alphanpi}$ and $M_{alphanpi}^S$ models. The latter shows programs for the $M_{alphaargs}$ and $M_{alphaargs}^S$ models. 
Table \ref{tab:quicksort-programs-pre-conditions} shows the programs preconditions. 

\begin{table*}[h!]
\centering
\caption[Hierarchy of the programs used to learn Quicksort.]{Hierarchy of the programs for Quicksort. In this case, the programs cannot accept arguments. The level 0 programs are not learned and they are the atomic actions.}
\resizebox{0.9\linewidth}{!}{%
\begin{tabular}{|l|p{0.7\linewidth}|l|}
\hline
\textbf{Program} & \textbf{Description} & \textbf{Level} \\ \hline
\texttt{quicksort} & Sort the list. & 5 \\ \hline
\texttt{quicksort\_update} & Execute \texttt{partition} call and populate the stack with the next pointers. & 4 \\ \hline
\texttt{partition} & Order a sub-array and return the pivot for the next operation. & 2 \\ \hline
\texttt{partition\_update} & Swap the elements pointed by $p_3$ (pivot) and $p_1$ if $p_3$ pointed value is less than $p_2$ pointed value. & 1 \\ \hline
\texttt{save\_ptr\_1} & Save pointer 1 position into the temporary registry. & 0 \\ \hline
\texttt{load\_ptr\_1} & Load the temporary registry value into the pointer 1 position. & 0 \\ \hline
\texttt{push} & Push pointers values inside the stack. & 0 \\ \hline
\texttt{pop} & Pop and restore pointers values from the stack. & 0 \\ \hline
\texttt{swap} & Swap elements pointed by pointer 1 and by pointer 2. & 0 \\ \hline
\texttt{swap\_pivot} & Swap elements pointed by pointer 1 and by pointer 3 (pivot). & 0 \\ \hline
\texttt{ptr\_i\_left} & Move pointer $i$ (where $i \in \{1,2,3\}$) one step left. & 0 \\ \hline
\texttt{ptr\_i\_right} & Move pointer $i$ (where $i \in \{1,2,3\}$) one step right. & 0 \\ \hline
\end{tabular}%
}
\label{tab:quicksort-programs}
\end{table*}

\begin{table*}[h!]
\centering
\caption[Hierarchy of the programs used to learn Quicksort.]{Hierarchy of the programs for Quicksort. In this case, the atomic actions can accept arguments. The level 0 programs are not learned and they are the atomic actions.}
\resizebox{0.9\linewidth}{!}{%
\begin{tabular}{|l|l|l|l|}
\hline
\textbf{\textbf{Program}} & \textbf{\textbf{Description}} & \textbf{Arguments} & \textbf{Level} \\ \hline
\texttt{quicksort} & Sort the list. & 0 & 5 \\ \hline
\texttt{quicksort\_update} & Execute \texttt{partition} call and populate the stack with the next pointers. & 0 & 4 \\ \hline
\texttt{partition} & Order a sub-array and return the pivot for the next operation. & 0 & 2 \\ \hline
\texttt{partition\_update} & Swap the elements pointed by $p_3$ (pivot) and $p_1$ if $p_3$ pointed value is less than $p_2$ pointed value. & 0 & 1 \\ \hline
\texttt{save\_ptr} & Save a given pointer position into the temporary registry. & 1 & 0 \\ \hline
\texttt{load\_ptr} & Load the temporary registry value into the given pointer position. & 1 & 0 \\ \hline
\texttt{push} & Push pointers values inside the stack. & 0 & 0 \\ \hline
\texttt{pop} & Pop and restore pointers values from the stack. & 0 & 0 \\ \hline
\texttt{swap} & Given two pointers, swap the corresponding elements & 2 & 0 \\ \hline
\texttt{ptr\_left} & Move pointer one, two or all three pointers one step left. & 3 & 0 \\ \hline
\texttt{ptr\_right} & Move pointer one, two or all three pointers one step right. & 3 & 0 \\ \hline
\end{tabular}%
}
\label{tab:quicksort-prog-args}
\end{table*}

\begin{table*}[h!]
\centering
\caption{Pre-conditions associated to each Quicksort program.}
\resizebox{0.9\linewidth}{!}{%
\begin{tabular}{|l|p{0.7\linewidth}|}
\hline
\textbf{Program} & \textbf{Pre-condition} \\ \hline
\texttt{quicksort} & Pointer 1 and 3 are at the extreme left of the list. Pointer 2 is at the extreme right of the list. The stack is empty. \\ \hline
\texttt{quicksort\_update} & The stack is not empty and the temporary registry is empty. \\ \hline
\texttt{partition} & The temporary registry must contain Pointer 1 position. \\ \hline
\texttt{partition\_update} & Pointer 1 position must be lower than Pointer 3 position. Pointer 3 position must be lower than Pointer 2 position. The temporary registry must be not empty. \\ \hline
\texttt{save\_ptr\_1} & No pre-condition. \\ \hline
\texttt{load\_ptr\_1} & The temporary stack must not be empty. \\ \hline
\texttt{push} & Check custom boolean condition $(p_1+1 < p_2) \vee (p_1-1 > 0 \wedge p_3 < p_1-1)$. \\ \hline
\texttt{pop} & The stack must not be empty. \\ \hline
\texttt{swap} & Pointer 1 position must be different than pointer 2 position. \\ \hline
\texttt{swap\_pivot} & Pointer 1 position must be different than pointer 3 position. \\ \hline
\texttt{ptr\_i\_left} & Pointer $i$ (where $i \in \{1,2,3\}$) is not at the extreme left of the list. \\ \hline
\texttt{ptr\_i\_right} & Pointer $i$ (where $i \in \{1,2,3\}$) is not at the extreme right of the list. \\ \hline
\end{tabular}%
}
\label{tab:quicksort-programs-pre-conditions}
\end{table*}

\end{document}